\documentclass[11pt]{article}
\usepackage{wrapfig}
\usepackage{moriond,graphicx}

\def\be{\begin{equation}}
\def\ee{\end{equation}}
\def\bea{\begin{eqnarray}}
\def\eea{\end{eqnarray}}

\begin{document}
\vspace*{4cm}
\title{Jet Measurements and Determination of $\alpha_S$ at HERA}

\author{ Ch. Wissing\,\footnote{on behalf of the H1 and ZEUS collaborations} }

\address{Institut f{\"u}r Physik \\ Universit{\"a}t Dortmund \\ 44221
  Dortmund, Germany \\ wissing@physik.uni-dortmund.de}

\maketitle\abstracts{
The strong coupling constant $\alpha_S$ can be measured in deep
inelastic $ep$ scattering by employing various methods. Recent results on
the extraction of $\alpha_S$ using jet and event shape observables are
presented. The results are found to be in good agreement with other
more inclusive measurements and the world average value.
}

\section{Introduction}

High energetic electron proton collisions provided by the HERA
accelerator allow precision tests of Quantum Chromo Dynamics (QCD).
The collected data spread over a wide range in negative four-momentum
 transfer $Q^2$ and transverse energy $E_T$ of the jets in the
 hadronic final state.
The jet production cross-section can be expressed as a
convolution of the hard scattering matrix
element $\hat{\sigma}_{QCD}$ with the proton parton
density functions (PDFs):
\begin{equation}
\label{e:JetCrossSection}
\sigma_{jet} = \sum_{i=q,\bar{q},g} \int dx ~
\mbox{pdf}_i(x,\mu_f,\alpha_S)
\otimes  \hat{\sigma}_{QCD}(x,\mu_r,\mu_f,\alpha_S)
\cdot (1+ \delta_{had}), 
\end{equation}
with $\mu_f$ being the factorisation scale, $\mu_r$ the
renormalisation scale and $\delta_{had}$ the correction from parton to
hadron level.

A suitable frame to study jet production in neutral current deep
inelastic scattering at HERA is the Breit frame
of reference, where in the Quark Parton Model (QPM) the struck quark
and the virtual boson collide head on. Any 
observation of significant transverse energy in the final state can be
attributed to higher order QCD processes.

In order to determine $\alpha_S$ a Next-to-leading order (NLO)
calculation is employed to provide the $\alpha_S$ dependence of the
predicted cross-section, \mbox{$\sigma^{theo}=A \cdot \alpha_S(M_Z)+B \cdot
\alpha_S^2(M_Z)$} with parameters $A$ and $B$ which are determined by
a fit. The measured cross-section is then used to obtain a value of
$\alpha_S(M_Z)$ by employing this parametrisation.

\section{Inclusive Jet Cross-Section}

\begin{wrapfigure}[24]{r}{0.5\textwidth}
\center
\includegraphics[width=0.43\textwidth,bb=45 0 460 566,clip=true]{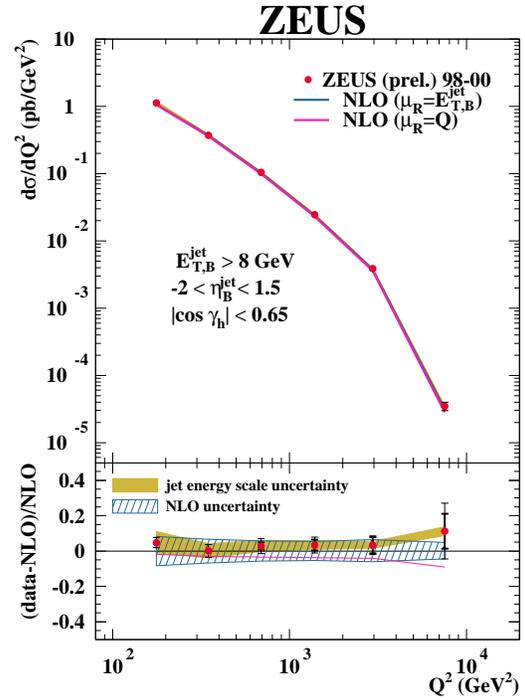}
\begin{minipage}{0.5\textwidth}
\sloppy
\caption{Inclusive jet cross-section as function of $Q^2$.}
\label{f:JetX_Zeus}
\end{minipage}
\end{wrapfigure}
The inclusive jet analysis by the ZEUS
 collaboriation\,\cite{r:IncJetsZeus}  used a data set of an integrated
luminosity of \mbox{$ 81.7\,\mbox{pb}^{-1}$}. Jets were found by the
longitudinally invariant $k_T$ algorithm applied in the Breit
frame. For the event selection at least one jet with
\mbox{$E_T^{Breit}>8\,\mbox{GeV}$} and \mbox{$Q^2>125\,\mbox{GeV}^2$} were
required. The cross-section has been measured single- and double
differentially in $Q^2$, $E_T$ and also the pseudo-rapidity $\eta$ of
the jets in the Breit frame. The cross-section as a function of $Q^2$
is compared to an NLO prediction based on the DISENT\,\cite{r:disent}
 package in Fig~\ref{f:JetX_Zeus}. 

The uncertainties of the prediction are \mbox{$\sim 7\,\%$}.  They
arise from missing 
higher orders in the calculation and limited precision of the proton PDFs.

The measured cross-section has been
corrected for detector effects and QED radiation by
using a leading order (LO) Monte Carlo generator. The experimental
uncertainties are dominated by the knowledge of the absolute
hadronic energy scale of the calorimeter.

A very precise value for $\alpha_S(M_Z)$ has been extracted for
\mbox{$Q^2>500\,\mbox{GeV}^2$}, where theoretical uncertainties are small:
$$
\alpha_S(M_Z) = 0.1196 \pm 0.0011 (\mbox{stat.}) ^{+0.0019}_{-0.0025}
(\mbox{exp.}) ^{+0.0029}_{-0.0017} (\mbox{theo.}).
$$

A very similar analysis\,\cite{r:IncJetsH1} has been carried out by H1
collaboration using a data set of 
\mbox{$61.25\,\mbox{pb}^{-1}$} of luminosity. The most important
selection criteria were \mbox{$E_T^{Breit}>7\,\mbox{GeV}$} and\\
\mbox{$150\,\mbox{GeV}^2 < Q^2 < 5000\,\mbox{GeV}^2$}. The NLO program
used in the extraction of $\alpha_S$ was
NLOJET++\,\cite{r:nlojet}. The obtained value for $\alpha_S$ is:
$$
\alpha_S(M_Z) = 0.1197 \pm 0.0016 (\mbox{exp.}) ^{+0.0046}_{-0.0048} (\mbox{theo.}).
$$

\section{Multi Jet Cross-Section}
A measurement of the three jet cross-section is well suited for a
determination of \mbox{$\alpha_S$} since the lowest order contribution to the
prediction for this
event class is proportional $\alpha_S^2$. Many uncertainties of the
measurement cancel out if the three to two jet ratio $R_{3/2}$ is
investigated.

A recent multi jet analysis\,\cite{r:MultiJetsH1} by the H1
Collaboration used a data set of
\mbox{$65.4\,\mbox{pb}^{-1}$} of integrated luminosity. The considered
kinematical domain was given by \mbox{$150 < Q^2 <
  15000\,\mbox{GeV}^2$} and \mbox{$E_T > 5\,\mbox{GeV}$}.
To ensure an infra-red safe region for the
NLO predictions an invariant jet mass $M_{3jet}>25\,\mbox{GeV}$
($M_{2jet}>25\,\mbox{GeV}$) was required for the three (two) jet sample.


The NLO prediction has been calculated by employing the NLOJET++
package. $Q^2$ has been chosen for the squared renormalisation scale
and it has been varied by a factor 4 to estimate the influence
of missing higher order contributions.

\begin{figure}[ht]
\center
\includegraphics[width=0.40\textwidth]{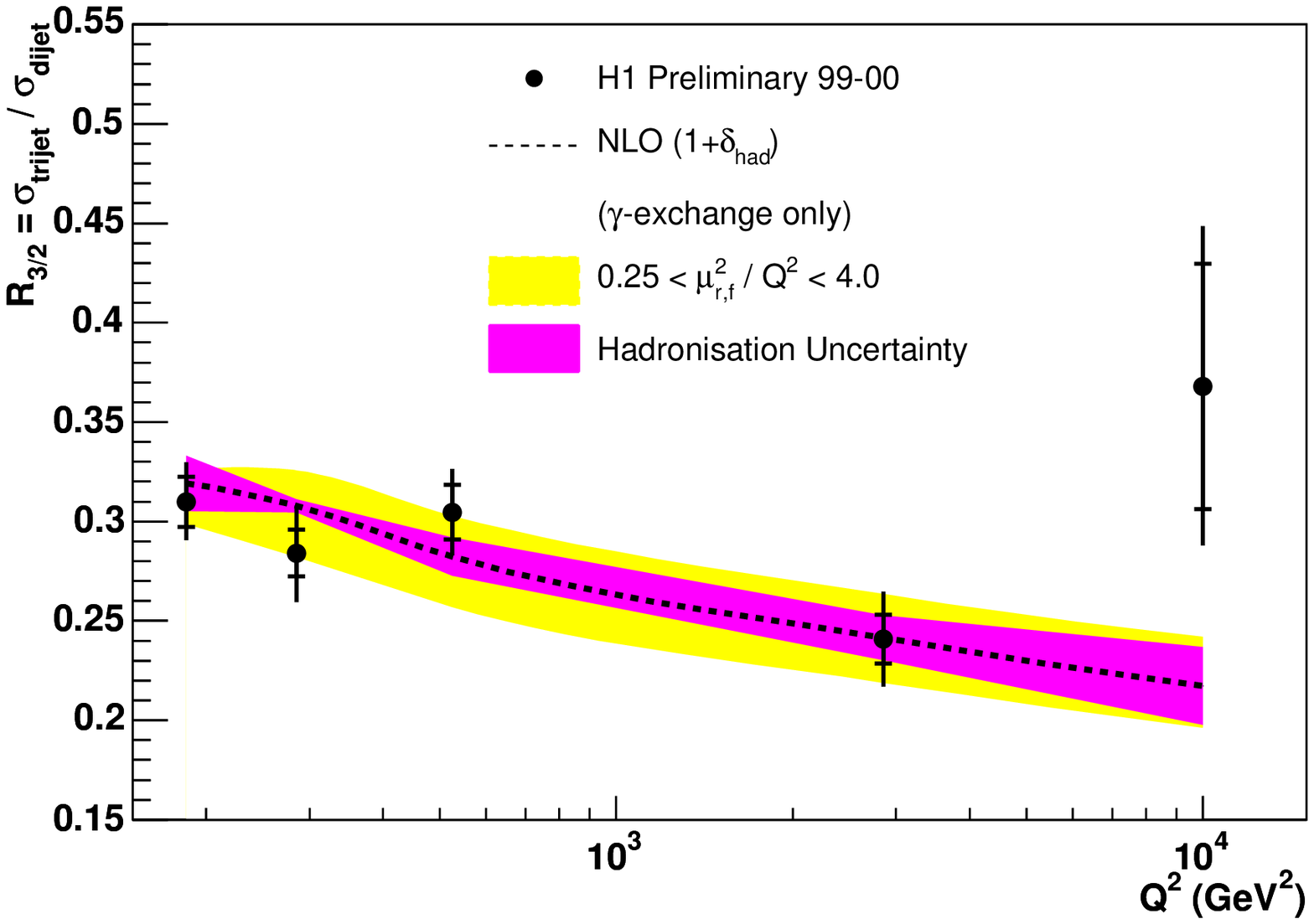}
\hspace{0.05\textwidth}
\includegraphics[width=0.40\textwidth]{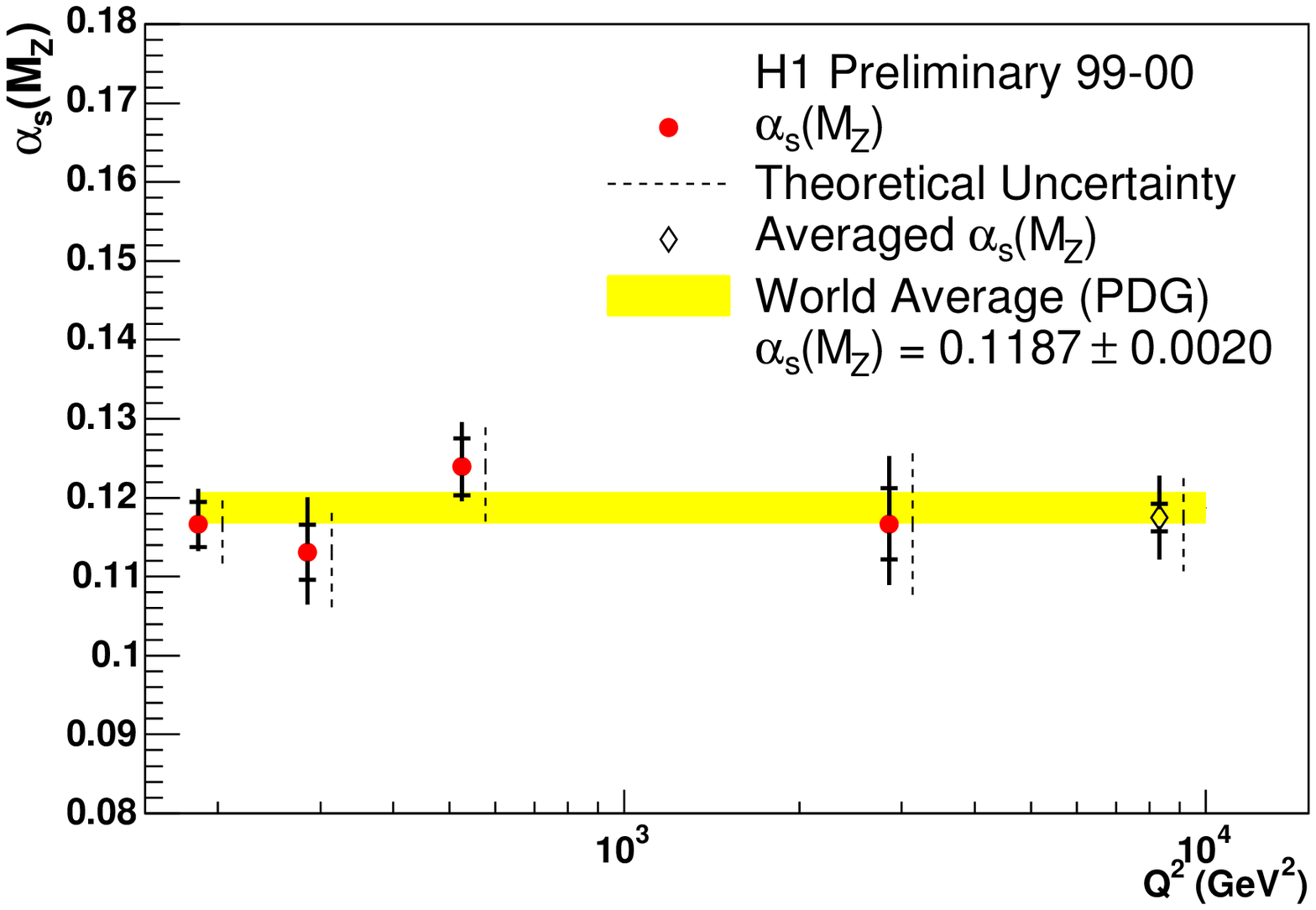}
\caption{$R_{3/2}$ as a function of
  $Q^2$ (left). Extracted $\alpha_S(M_Z)$ values compared to the world
  average (right).}
\label{f:MultiJetsH1}
\end{figure}

Fig\,\ref{f:MultiJetsH1}\,(left) shows $R_{3/2}$ as a function of $Q^2$. As
there are no electroweak effects included in the NLO prediction the
observed deviation in the highest $Q^2$ bin is expected and the bin
has been excluded therefore in the $\alpha_S$ extraction.

The $\alpha_S(M_Z)$ values for the remaining $Q^2$ bins are displayed in
Fig\,\ref{f:MultiJetsH1}\,(right) together with an averaged value, which
is found to be
$$
\alpha_S(M_Z) = 0.1175 \pm 0.0017 (\mbox{stat.}) \pm 0.0050
(\mbox{exp.}) ^{+0.0054}_{-0.0068} (\mbox{theo.}).
$$
This result and the ones presented in section~2 are in good agreement
with the world average.

\section{Event Shapes}
In contrast to the jet measurements
the method of event
shapes considers characteristics of the complete final state.
To avoid non-perturbative effects due to the proton remnant the event
shape variables are usually defined in the current hemisphere of the
Breit frame.

As an alternative approach to phenomenological models for the
hadronisation a power correction technique has been developed\,\cite{r:PC}. Mean
values (or differential distributions) get shifted by a constant:
$$
\langle F \rangle = \langle F \rangle^{QCD} + a_F \cal{P}
$$
where $a_F$ is of order one and calculable perturbatively. The power
correction term $\cal{P}$ is proportional to $1/Q$ and depends on a universal
parameter $\alpha_0$ and $\alpha_S$.

\begin{figure}[hb]
\center
\includegraphics[width=0.35\textwidth, bb=0 0 580 520,clip=true]{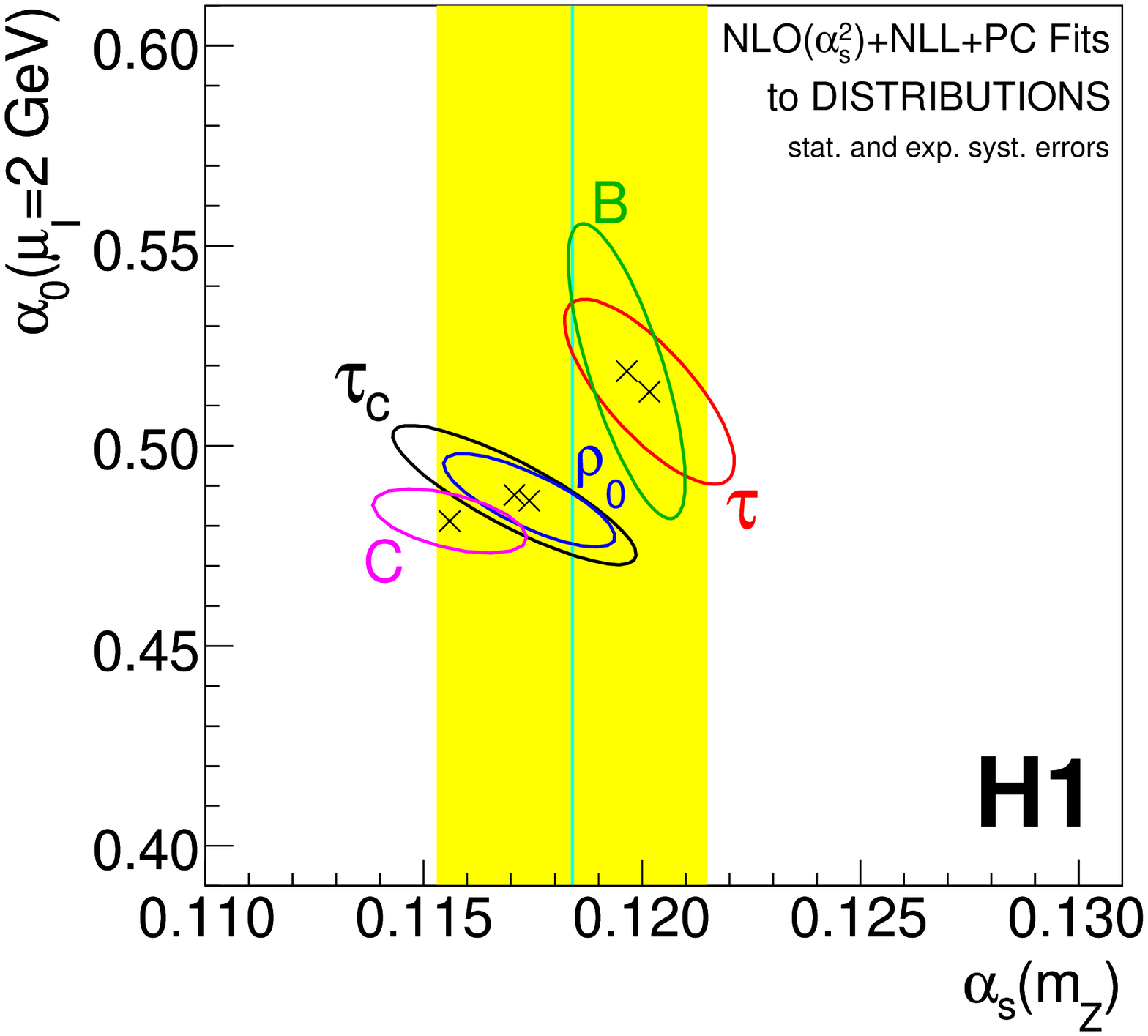}
\hspace{0.1\textwidth}
\includegraphics[width=0.33\textwidth]{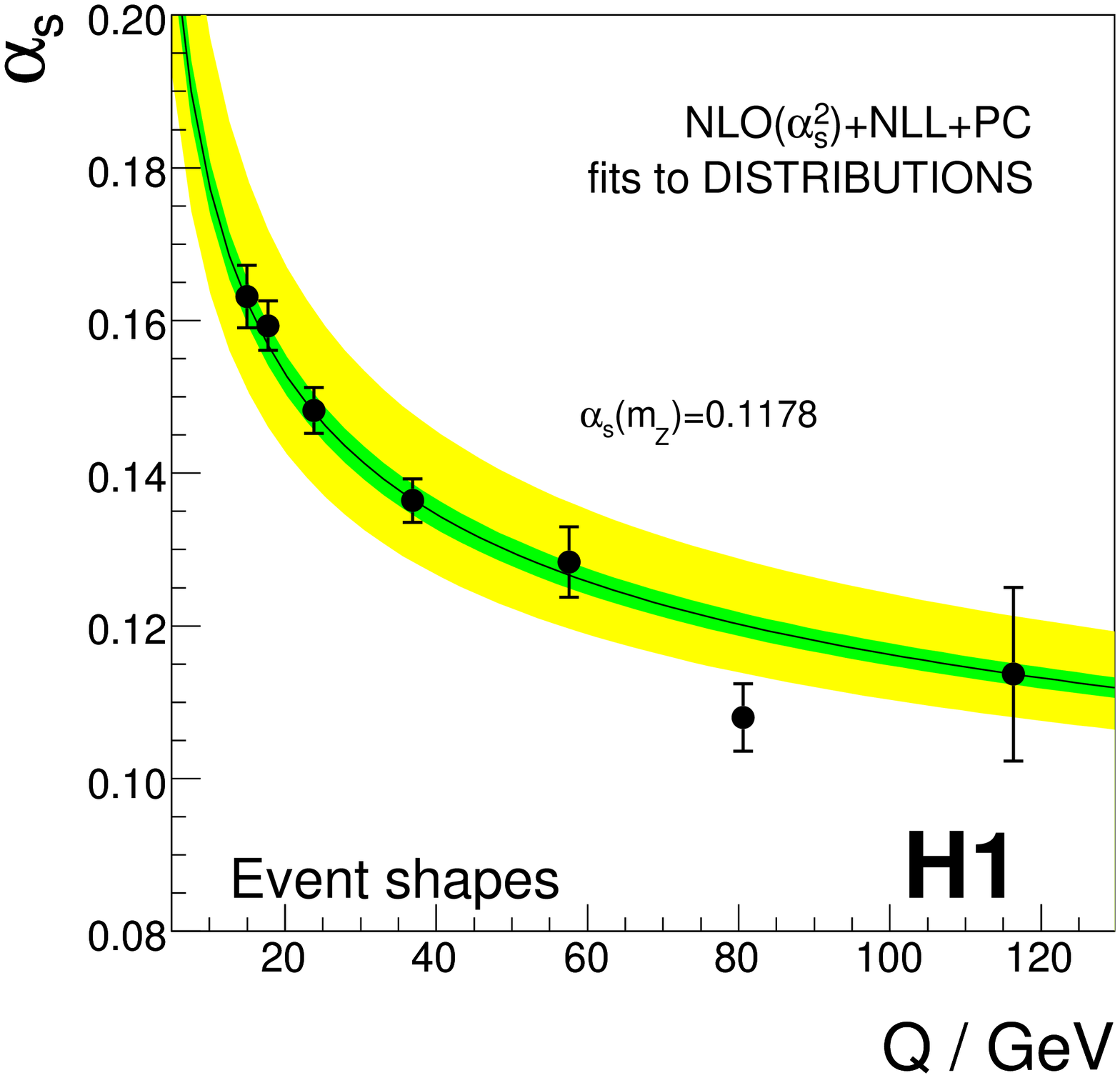}
\caption{Fits to various differential event shape distributions in the
  $(\alpha_S,\alpha_0)$ plane (left). Running $\alpha_S(Q)$ (right).}
\label{f:EventShapes}
\end{figure}

A recent H1 analysis\,\cite{r:EventShapesH1} observed for event shape
distributions a good agreement
between the data points and fits based on an NLO-QCD calculation
including a resummation supplemented by power
corrections\,\cite{r:pc}. A $\chi^2$ 
fit to these distributions allowed a simultaneous determination of
$\alpha_0$ and
$\alpha_S$. The result for various event shape variables is shown in
Fig\,\ref{f:EventShapes}\,(left). The results give a consistent value
for $\alpha_S$
and support a universal $\alpha_0$ parameter of about 0.5 as it is
expected by theory. An averaging procedure gives:
\begin{eqnarray*}
\alpha_S(M_Z) & = & 0.1198 \pm 0.0013 (\mbox{exp}) ^{+0.0056}_{-0.0043} (\mbox{theo.})\\
     \alpha_0 & = & 0.476 \pm 0.008 (\mbox{exp}) ^{+0.018}_{-0.059} (\mbox{theo.})\,.
\end{eqnarray*}

In a slightly modified fit procedure values of $\alpha_S(Q)$ can be
derived. The running of the strong coupling constant can be nicely
observed in Fig\,\ref{f:EventShapes}\,(right).


\section{Summary}

\begin{wrapfigure}[16]{r}{0.5\textwidth}
\center
\vspace{-0.7cm}
\includegraphics[width=0.42\textwidth,bb=40 0 500 520,clip=true]{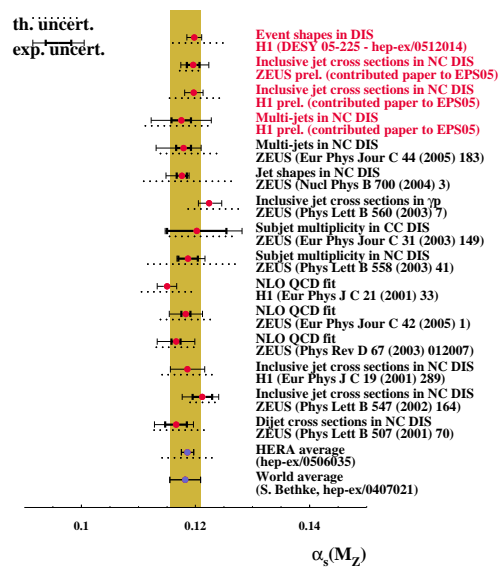}
\caption{Comparison of $\alpha_S$ extractions from HERA.}
\label{f:AllAlphaS_hera}
\end{wrapfigure}
The strong coupling constant $\alpha_S$ has been extracted from
various measurements performed by the H1 and ZEUS collaborations. The
results are found to be well compatible with each other 
independent of the method and in excellent agreement with the world
average. Fig\,\ref{f:AllAlphaS_hera} shows a comparison of the
presented $\alpha_S$ extractions to a combined HERA
value\,\cite{r:Claudia} and the world average, which up to now does
not consider the HERA measurements.

With increased data sets presently collected at HERA-II the
statistical and experimental uncertainties will be further
reduced. Calculations including NNLO QCD contributions and electroweak
effects are needed to reduce the theoretical uncertainties.


\section*{Acknowledgements}
In particular I want to thank my colleagues Claudia Glasman and
Thomas Kluge for their help to prepare the presentation and the
proceedings.
Financial support was provided by the European Union and
the German Bundesministerium f{\"u}r Bildung und Forschung (05\,H14PEA/6).

\enlargethispage{16pt}

\end{document}